\documentclass[journal]{IEEEtran}
\usepackage[dvips]{graphicx}
\graphicspath{{fig/}}
\usepackage{url}
\usepackage{graphicx}
\usepackage{epstopdf}
\usepackage{booktabs}
\usepackage{threeparttable}
\usepackage{multirow}
\usepackage{color}
\usepackage{hyperref}
\usepackage{balance}
\usepackage{amsmath}
\usepackage{amssymb}

\begin{document}

\title{An Olfactory EEG Signal Classification Network Based on Frequency Band Feature Extraction}

\author{Biao Sun$^{1}$, \IEEEmembership{Senior Member, IEEE}, Zhigang Wei$^{1}$, Pei Liang$^{2}$, and Huirang Hou$^{1,*}$, \IEEEmembership{Member, IEEE}

\thanks{This work was supported by the National Natural Science Foundation of China under Grant 61971303 and the China Postdoctoral Science Foundation 2021M692390.}

\thanks{$^{1}$ School of Electrical and Information Engineering, Tianjin University, Tianjin 300072, China.}

\thanks{$^{2}$ College of Optical and Electronic Technology, China Jiliang University, Hangzhou 310018 China.}

\thanks{$^{*}$ Corresponding author: Huirang Hou. Email: houhuirang@tju.edu.cn.}

\thanks{The authors would like to thank Professor Qinghao Meng from School of Electrical and Information Engineering of Tianjin University for sharing the olfactory EEG dataset.}

}

\maketitle

\begin{abstract}
Classification of olfactory-induced electroencephalogram (EEG) signals has shown great potential in many fields. Since different frequency bands within the EEG signals contain different information, extracting specific frequency bands for classification performance is important. Moreover, due to the large inter-subject variability of the EEG signals, extracting frequency bands with subject-specific information rather than general information is crucial. Considering these, the focus of this letter is to classify the olfactory EEG signals by exploiting the spectral-domain information of specific frequency bands. In this letter, we present an olfactory EEG signal classification network based on frequency band feature extraction. A frequency band generator is first designed to extract frequency bands via the sliding window technique. Then, a frequency band attention mechanism is proposed to optimize frequency bands for a specific subject adaptively. Last, a convolutional neural network (CNN) is constructed to extract the spatio-spectral information and predict the EEG category. Comparison experiment results reveal that the proposed method outperforms a series of baseline methods in terms of both classification quality and inter-subject robustness. Ablation experiment results demonstrate the effectiveness of each component of the proposed method.
\end{abstract}


\IEEEpeerreviewmaketitle

\section{Introduction}

\IEEEPARstart{E}{lectroencephalogram} (EEG) signals are an important source of information for studying brain activity, and its current research is mainly focused on visual \cite{petrantonakis2012adaptive}, auditory \cite{biesmans2016auditory}, and motor imagery \cite{sun2021adaptive}. As an essential sense, the olfactory can directly stimulate memory and trigger strong emotions \cite{ishida2002improvement}. Due to the great potential of the olfactory, there have been substantial studies devoted to exploring olfactory EEG signals. Previous research has shown that the classification of olfactory EEG signals plays a vital role in many fields, including neuroscience research \cite{lin2018sniffing}, disorder treatment \cite{saha2014eeg}, multimedia \cite{kroupi2015subject}, and brain-computer interfaces \cite{placidi2015basis}.

For EEG signal analysis, feature extraction is of particular importance. To date, various methods of extracting features have been used in EEG signal analysis \cite{hou2017improving, hou2020olfactory, ezzatdoost2020decoding}. Power-spectral-density (PSD) is a popular method for feature extraction, which explores spectral-domain information in EEG signals \cite{lanata2016automatic}. Since different frequency bands of the EEG signals contain different information, many methods have been proposed to extract specific frequency bands for EEG signal classification. In the common spatial pattern-based research \cite{meng2014simultaneously, liu2015boosting, das2016discriminative}, to identify specific frequency bands, \cite{ang2008filter} exploits the mutual information of the frequency bands, and \cite{higashi2012simultaneous} exploits the Fisher ratio of frequency band power.

In the traditional PSD-based EEG signal classification approach, the EEG frequency spectrum is divided into five sub-bands based on biological rhythms: $\delta$ (0.5-4 Hz), $\theta$ (4-7.5 Hz), $\alpha$ (8-13 Hz), $\beta$ (13-30 Hz), $\gamma$ ($>$30 Hz), and the PSD is extracted from these sub-bands \cite{fang2020multi, frantzidis2010toward}. Recently an average frequency band division (AFBD) method is proposed, which averagely divides the PSD curve into several sub-bands with equal bandwidth and obtains significant breakthroughs for frequency band division \cite{hou2020odor}. These methods of extracting frequency bands are heuristic or manual, leading the extracted frequency bands to be shared across subjects. However, EEG signals have large inter-subject variability, and the shared across subjects frequency bands are general, ignoring informative differences between subjects. So the existing methods of extracting frequency bands can be improved. \\
\indent Recent advances in deep learning technologies have exhibited promising capabilities in handling inter-subject variability. In \cite{zhang2021motor}, the deep learning technology is used in EEG channel selection aimed to generate the subject-specific channel combination. In \cite{zhang2019convolutional}, the authors utilize the deep learning technology to explore the most discriminative temporal periods for each subject. In \cite{fang2020learning}, based on deep learning technology, a novel regional attention CNN is designed to adaptively identify the activated area of the primary sensorimotor for each subject. While there are currently few explorations on frequency band extraction using deep learning technology, its potential for extracting frequency bands has only begun to be explored.\\
\indent In this letter, we present an olfactory EEG signal classification network (OESCN) based on frequency band feature extraction, which aimed to exploit the characteristics of EEG signals in spectral-domain to classify olfactory EEG signals. First, a frequency band generator is designed to extract frequency bands based on PSD features via the sliding window technique. Then, a frequency band attention mechanism is proposed to reweight frequency bands for a specific subject. Finally, a CNN is constructed to extract the spatio-spectral information of the encoded EEG signals and predict the EEG category. Comparison experiment results reveal that the proposed method outperforms a series of baseline methods. Ablation experiment results demonstrate the effectiveness of each component of the proposed method.

\begin{figure*}[htbp]
	\centering
	\includegraphics[width=1\textwidth]{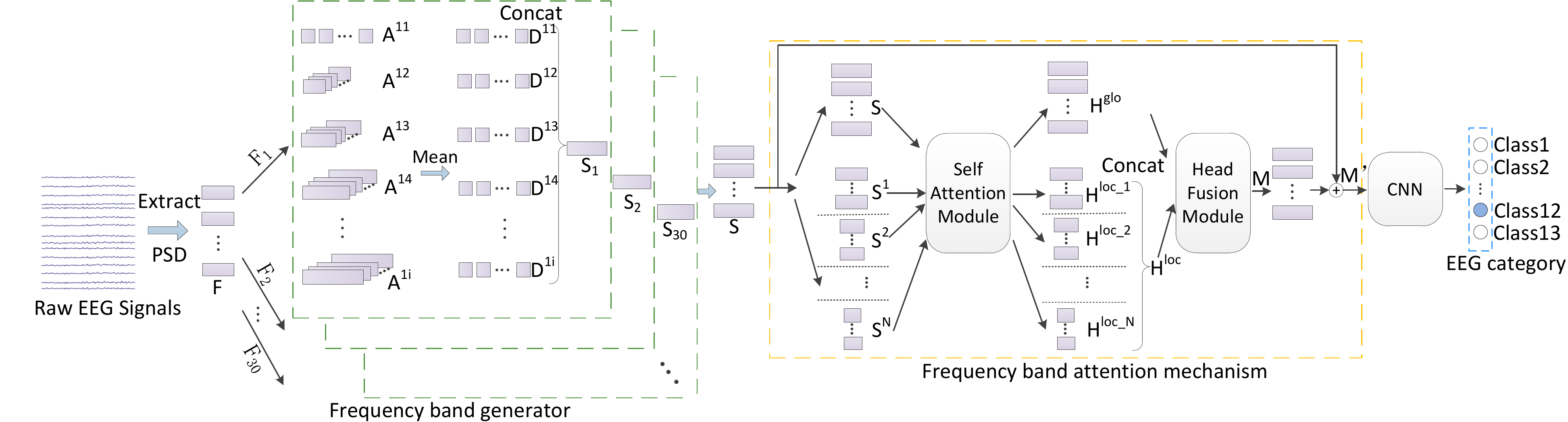}
	\caption{Overview of the proposed olfactory EEG signal classification network based on frequency band feature extraction. A Welch periodogram is firstly applied to estimate PSD features from the raw EEG signals. Secondly, a frequency band generator is designed to extract frequency bands based on PSD features using the sliding window technique. Thirdly, a frequency band attention mechanism adaptively optimizes frequency bands for a specific subject. Lastly a CNN is constructed as the classifier to predict the EEG category.}
	\label{fig:Fig. 1}
\end{figure*}

\section{METHOD}

\subsection{Pipeline Overview}

Fig. \ref{fig:Fig. 1} shows an overview of the proposed OESCN. The goal is to predict the EEG category based on the raw EEG signals $\textbf{X} \in \mathbb{R} ^{C \times T}$ with $C$ EEG channels and $T$ time points. For each channel, a Welch periodogram is firstly applied to estimate PSD features for frequencies between 0.5 Hz and 70 Hz. Secondly, a designed frequency band generator uses the sliding window technique to extract frequency bands containing different EEG information from PSD features. Thirdly, a frequency band attention mechanism is proposed to reweight frequency bands to optimize for a specific subject. Lastly, a CNN is constructed as the classifier to predict the EEG category.

\subsection{Frequency Band Generator}

In the frequency band generator, as shown on the left side of Fig. \ref{fig:Fig. 1}, a Welch periodogram is first applied to the EEG signals $\textbf{X}$ at each channel to estimate PSD features $\textbf{F} \in \mathbb{R}^{C \times P}$ between 0.5 Hz and 70 Hz, where $P$ is the number of PSD features. The Hamming window of the Welch periodogram is set to 200, and the number of overlap points between adjacent windows of the Welch periodogram is set to 8.

 For each channel of the PSD features $\textbf{F}_c  \in \mathbb{R}^{1 \times P}$, a sliding window technique splits $\textbf{F}_c$ into several frequency band combinations $\{{\textbf{A}^{ci} \in \mathbb{R}^{1 \times L_{ci} \times{B_{ci}}},i = 1,...,N}\}$, where $i$ is the index of frequency band combinations, $N$ is the number of frequency band combinations, ${L_{ci}}$ is the window length of a frequency band slice and ${B_{ci}}$ is the number of frequency bands. The ${B_{ci}}$ is calculated by:
\begin{equation}
{B_{ci}} = \left\lfloor {(P - {L_{ci}})/G} \right\rfloor ,
\end{equation}
where $G$ is the increment between two neighboring frequency band slices. Then, the averaging method is used to calculate the mean of each frequency band for each ${\textbf{A}^{ci}}$, the new frequency band combinations are $\{{\textbf{D}^{ci} \in \mathbb{R}^{1 \times{B_{ci}}},i = 1,...,N}\}$. Then, all ${\textbf{D}^{ci}}$ are concatenated along the channel dimension into a frequency band combination $\left\{ {\textbf{S}_c \in {\mathbb{R}^{1 \times K}}\left| {\textbf{S}_c = {\rm{Concat}}({\textbf{D}^{ci}}),i = 1,...,N} \right.} \right\}$, where $K$ is calculated by:
\begin{equation}
K = \sum\nolimits_i {{B_{ci}}}.
\end{equation}
Lastly, all ${\textbf{S}_c}$ are concatenated along the frequency band combination dimension into a frequency band combination $\textbf{S}\in {\mathbb{R}^{C \times K}}$.

In this work, a frequency band generator is designed to extract the frequency band combination $\textbf{S}$  based on the PSD features $\textbf{F}$. Since different frequency bands of the EEG signal may contain different information, our goal is to extract as many useful frequency bands as possible to form a frequency band combination. Inspired by the traditional EEG frequency spectrum division method based on biological rhythm involving five sub-bands \cite{becerra2018odor} and the recent average frequency band division method in which the PSD features are divided into several sub-bands with the same frequency bandwidth \cite{hou2020odor}, we select five window lengths of frequency band slice: 1, 5, 10, 15, and 20 Hz, i.e., ${{L_i} \in \left\{ {1, 5, 10, 15, 20} \right\}}$. The increment between two neighboring frequency band slices $G$ is consistently 1 Hz.
\begin{figure}[htbp]
	\centering
    \includegraphics[width=0.46\textwidth]{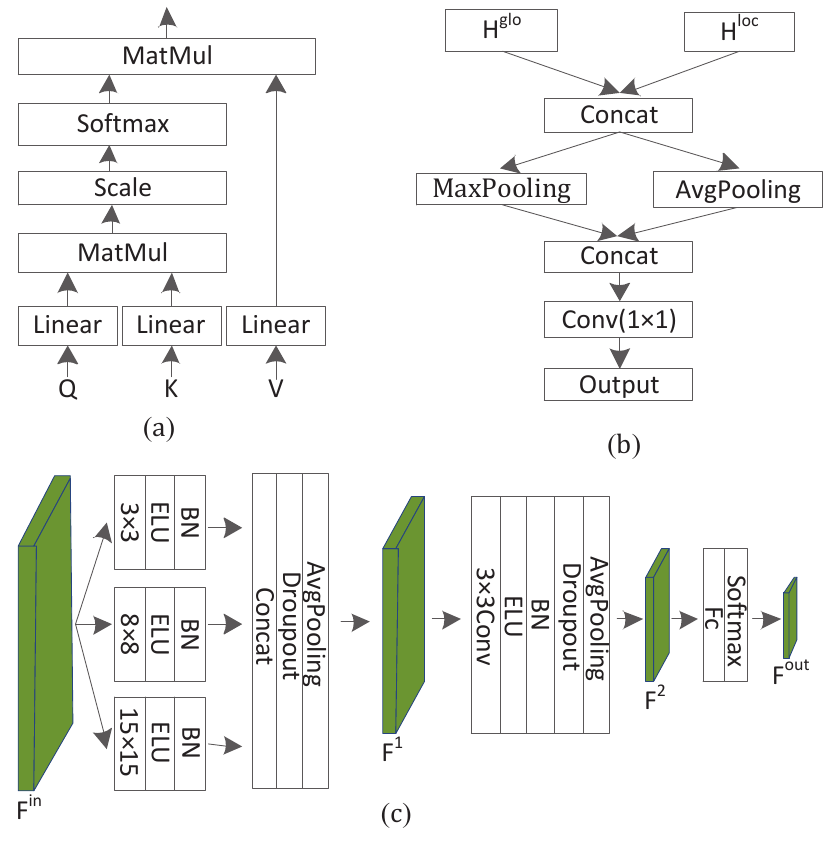}
	\caption{Schematic diagrams of the (a) self-attention module, (b) head fusion module, and (c) CNN of the proposed OESCN.}
	\label{fig:Fig 2}
\end{figure}

\subsection{Frequency Band Attention Mechanism}
As shown on the right side of Fig. \ref{fig:Fig. 1}, inspired by the multi-head attention mechanism \cite{vaswani2017attention}, we design a frequency band attention mechanism to reweight frequency bands, optimizing the frequency band combination suitable for a specific subject. The frequency band combination $\textbf{S}$ is calculated in parallel by 1+N heads of frequency band attention mechanism, which allows the model to attend information from the global and local subspaces.

In the first head, $\textbf{S}$ as a whole is used as the input of the self-attention module to utilize the global information between frequency bands of all bandwidths. In the self-attention module [Fig. \ref{fig:Fig 2} (a)], $\textbf{S}$ is first linearly transformed into three latent spaces:
\begin{equation}
{\textbf{Q}^{\rm{glo}} = }\textbf{S}{\textbf{W}^{\rm{query}}},  \quad \, {\textbf{K}^{\rm{glo}} = }\textbf{S}{\textbf{W}^{\rm{key}}},  \quad {\textbf{V}^{\rm{glo}} = }\textbf{S}{\textbf{W}^{\rm{value}}},
\end{equation}
where $\textbf{Q}^{\rm{glo}}$, $\textbf{K}^{\rm{glo}}$, $\textbf{V}^{\rm{glo}}\in { \mathbb{R}^{C \times K}}$, and ${\textbf{W}^{\rm{query}}}$, ${\textbf{W}^{\rm{key}}}$, ${\textbf{W}^{\rm{value}}}$ are parameter matrices  $ \in { \mathbb{R}^{K \times K}}$. We compute the dot products of the $\textbf{Q}^{\rm{glo}}$ with $\textbf{K}^{\rm{glo}}$, divide each by $C$, where $C$ is the number of EEG electrode nodes. Finally, we apply a softmax function to obtain the weights on $\textbf{V}^{\rm{glo}}$. The output of the first head is:
\begin{equation}
{\textbf{H}^{\rm{glo}}} = \textbf{V}^{\rm{glo}}{\rm{Softmax}} (\frac{{{{\textbf{Q}^{\rm{glo}}}^T}\textbf{K}^{\rm{glo}}}}{\sqrt{C}}), \quad {\textbf{H}^{\rm{glo}}} \in { \mathbb{R}^{C \times K}}.
\end{equation}

For the rest of N heads, $\textbf{S}$ is firstly divided into multiple frequency band combinations ${{\textbf{S}^i} \in { \mathbb{R} ^{C \times B_{ci}} }}$ with the same bandwidth: $\left\{ {{\textbf{S}^i_c} = {\textbf{D}^{ci}},i = 1,...,N} \right\}$. Then each ${{\textbf{S}^i} }$ is served as the input of one head of N heads. In this way, local information between frequency bands of the same bandwidth is used. In each head, ${{\textbf{S}^i}}$ is calculated by self-attention module, the output of each head is ${\textbf{H}^{\rm{loc\_i}} \in { \mathbb{R} ^{C \times B_{ci}} }}$. Lastly, all ${\textbf{H}^{loc\_i}}$ are concatenated along the channel dimension:
\begin{equation}
{\textbf{H}^{\rm{loc}}} = {\rm Concat}( {{\textbf{H}^{{\rm loc\_i}}}}),\quad   {\textbf{H}^{\rm{loc}}} \in { \mathbb{R} ^{C \times K}}.
\end{equation}

To effectively fuse the outputs of I+N heads, we design a head fusion module as shown in Fig. \ref{fig:Fig 2} (b). The max-pooling layer and the average pooling layer can extract different information. The convolutional layer with a kernel size of (1,1) aggregates features into the same width as the input. The output of the head fusion module is:
\begin{equation}
\textbf{M} = {\rm{Conv}}\left\{ \begin{array}{l}
{\rm{Concat}}({\rm{MaxPooling}}\left[ {{\rm{Concat}}({\textbf{H}^{\rm{glo}}};{\textbf{H}^{\rm{loc}}}   })\right]\\;
  {\rm{AvePooling}}\left[ {{\rm{Concat}}({\textbf{H}^{\rm{glo}}};{\textbf{H}^{\rm{loc}}}   })  \right])
\end{array} \right\},
\end{equation}
where ${\textbf{M}} \in { \mathbb{R}^{C \times K}}$. We add the skip connection, the optimal frequency band combination $\textbf{M}'$ is calculated by:
\begin{equation}
{\textbf{M}'}{\rm{ = }} \textbf{M} + \textbf{S} , \quad   {\textbf{M}'} \in { \mathbb{R}^{C \times K}}.
\end{equation}

\subsection{classifier}
A specially designed CNN is constructed as the classifier to predict the EEG category [Fig. \ref{fig:Fig 2} (c)]. To extract the spatio-spectral information of $\textbf{M}'$, including the spatial information in EEG electrode nodes and the spectral information in frequency bands, the kernel size of convolutional layers are set as (n,n). The input $\textbf{M}'$ parallel goes through three convolutional layers of different kernel sizes (3$ \times $3, 8$ \times $8, 15$ \times $15) to extract features of different dimensions, followed by an ELU activation layer to maintain non-linearity. Then, the three outputs are concatenated along the channel dimension as the input of the average pooling layer to reduce the number of parameters and to extract important information. Then, the output goes through the convolutional layer with kernel size (3$ \times $3), the ELU activation layer, and the average pooling layer to further extra features. Lastly, three fully connected layers and the softmax activation function are used to predict the EEG category. In addition, the batch normalization (BN) layer and dropout layer with 0.25 keep probability are included to prevent overfitting.

\section{EXPERIMENTS AND RESULTS}

\begin{figure}[t]
	\centering
    \includegraphics[width=0.45\textwidth]{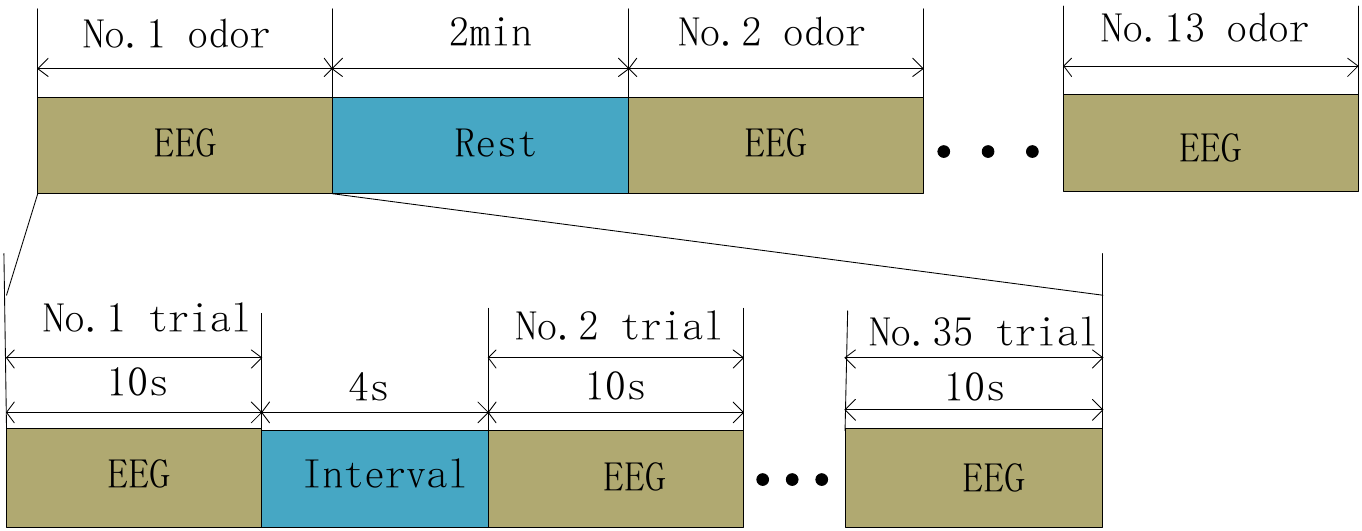}
	\caption{Schedule for a subject to conduct an experimental record.}
	\label{fig:Fig. 3}
\end{figure}

In this study, 13 odors were used for stimulation, including five olfactometer odors (rose, caramel, rotten smell, canned peaches, and excrement) and eight essential oils (mint, tea, coffee, rosemary, jasmine, lemon, vanilla, and lavender). Eleven healthy, right-handed subjects (eight males and three females), aged 24.9$\pm $3.0 years were recruited from the School of Electrical and Information Engineering at Tianjin University. They reported never having smoked or abused drugs and having no history of the respiratory, neurological, or olfactory disease. The experiment was conducted in a room of length 4 m, width 2 m, and height 2.5 m, which was equipped with an air conditioner to maintain the indoor temperature as well as an exhaust fan and an air purifier to eliminate residual indoor odor. The EEG signals were acquired using the Cerebus system at a sampling rate of 1 kHz. The 32 channels of the EEG cap were arranged according to the international 10-20 system. Of the 32 electrodes, all but two reference electrodes can be used for signal analysis. All EEG recording procedures were approved by the Tianjin University Ethics Committee
(No. TJUE-2020-186).

As shown in Fig. \ref{fig:Fig. 3}, for each subject, 13 classes of EEG signals were collected, corresponding to 13 kinds of odors. For each class of EEG signals, there were 35 trials. Each trial lasted 10s, and the interval between two trials was set to 4 seconds to avoid olfactory fatigue. Moreover, the rest time between two classes of odor stimulation was 2 minutes. Therefore, there were 455 (455 = 13 $\times$ 35) trials for each subject. Thus, there were 455 $\times$ 11 = 5005 trials in total.

We built and tested the FBEN using Pytorch in a Python 3.7 environment on a personal computer equipped with a 3.60-GHz Intel Core I7 central processor, an NVIDIA Titan Xp graphics processor, and 32 Gb of RAM. Data from each subject were divided into training and validation sets, and method performance was evaluated in terms of the classification accuracy metric. In every subject, trials were randomly divided into ten equivalent parts (10-fold cross-validation), one of which was retained as the validation set to test the method performance, while the other nine were used as the training set. During the 10-fold cross-validation, each of the ten parts served once as the validation set. The ten results were averaged together to produce a single estimate. Moreover, since the OESCN is applied to the olfactory EEG signal classification task, the cross-entropy loss function was used. The Adam optimizer with a learning rate of 0.0001 was used to train the network. The epoch number was 500, and the batch size was 39.

To evaluate the performance of the proposed OESCN, we benchmarked it against the EEGNet \cite{lawhern2018eegnet} and the AFBD-SVM method \cite{hou2020odor}. EEGNet is a classic deep learning network with excellent performance for processing the EEG signals, which encapsulated well-known EEG feature extraction concepts for brain-computer-interface in order to construct a uniform approach for different brain-computer interface paradigms. AFBD-SVM proposed a manual frequency band extraction method with excellent performance cooperating SVM as the classifier.

\begin{table}[t]
\footnotesize
    \begin{threeparttable}
		\caption{Classification accuracies of EEGNet, AFBD-SVM, OESCN, OESCN\_a1 AND OESCN\_a2}
        \label{table:label1}
        \setlength{\tabcolsep}{1.7pt}
        \begin{tabular}{lccccc}
        \specialrule{0em}{1pt}{1pt}
         \midrule[1pt]
            \multirow{2}{*}{Subject}&
            \multicolumn{5}{c}{Accuracy \% (mean $\pm$ std) }\cr
            \cmidrule(lr){2-6} & EEGNet & AFBD-SVM & OESCN & OESCN\_a1 & OESCN\_a2 \cr
            \midrule
        \;\; 1 & \textbf{87.1$\pm$4.9}  & 66.2$\pm$4.1          & 86.5$\pm$4.7              & 83.1$\pm$2.7   & 79.7$\pm$5.5\cr
        \;\; 2 & 61.3$\pm$4.3  & 90.1$\pm$3.8          & \textbf{97.8$\pm$1.8}              & 96.2$\pm$2.5   & 95.7$\pm$2.8\cr
        \;\; 3 & 95.8$\pm$2.6  & 97.8$\pm$1.4          & \textbf{99.6$\pm$0.9}              & 99.3$\pm$1.1   & 98.8$\pm$1.9\cr
        \;\; 4 & 65.8$\pm$9.9  & 70.8$\pm$6.1          & \textbf{99.2$\pm$1.3}              & 97.8$\pm$1.5   & 96.1$\pm$3.4\cr
        \;\; 5 & 88.3$\pm$3.3  & 85.3$\pm$5.8          & \textbf{97.2$\pm$2.3}              & 95.8$\pm$2.9   & 94.0$\pm$3.8\cr
        \;\; 6 & 81.9$\pm$8.0  & 91.9$\pm$3.1          & \textbf{99.0$\pm$1.0}              & 98.3$\pm$1.2   & 96.9$\pm$3.2\cr
        \;\; 7 & 89.4$\pm$4.4  & 89.5$\pm$3.9          & \textbf{95.4$\pm$1.2}              & 94.4$\pm$2.7   & 93.1$\pm$3.6\cr
        \;\; 8 & 97.2$\pm$2.5  & 98.5$\pm$1.0          & 99.8$\pm$0.9                 &\textbf{ 100$\pm$0}   & 99.3$\pm$1.1\cr
        \;\; 9 & 89.0$\pm$5.7  & 89.7$\pm$3.4          & \textbf{98.5$\pm$1.8}              & 96.5$\pm$2.8   & 93.3$\pm$2.6\cr
        \;\; 10 & 68.9$\pm$5.5 & 81.8$\pm$4.1          & \textbf{94.6$\pm$2.8}              & 93.7$\pm$3.2   & 91.2$\pm$2.4\cr
        \;\; 11 & 97.4$\pm$2.2 & 98.3$\pm$1.3          & \textbf{100$\pm$0}           &\textbf{ 100$\pm$0}   & 99.6$\pm$0.9\cr
        \midrule
        Average & 83.8 & 87.3       & \textbf{97.1}  & 95.9 & 94.3\cr
        \midrule
        Inter-subject std  & 12.3 & 10.2 & \textbf{3.7} & 4.5  & 5.3\cr
        \bottomrule[1pt]
      \end{tabular}
    \end{threeparttable}
\end{table}

As shown in Table \ref{table:label1}, we observe that the OESCN outperforms the other two methods, showing an average accuracy of 97.1${\rm{\% }}$, compared to 83.8${\rm{\% }}$ for EEGNet and 87.3${\rm{\% }}$ for AFBD-SVM. The results indicate that OESCN provides a 13.3${\rm{\% }}$ improvement with respect to EEGNet and a 9.8${\rm{\% }}$ improvement with respect to AFBD-SVM in terms of average accuracy. The inter-subject average standard deviations (std) of accuracy for three methods are shown in Table \ref{table:label1}. We observe that the inter-subject average standard deviation of OESCN (std = 3.7) is further lower than that of EEGNet (std = 12.3) and that of AFBD-SVM (std = 10.2), indicating the robustness to inter-subject variation of the proposed method. The intra-subject standard deviations are also shown in Table \ref{table:label1}. We observe that the intra-subject std of OESCN is lower than that of EEGNet and AFBD-SVM.

Next, an ablation study is then performed to evaluate the effectiveness of each component of the proposed OESCN. The necessity of the frequency band attention mechanism is first studied, we get a new network for performance comparison by removing the frequency band attention mechanism module from OESCN, which named OESCN\_a1. The necessity of the frequency band combination generator is then studied, we get a new network for performance comparison by removing the frequency band combination generator from OESCN\_a1, which named OESCN\_a2. The detailed comparison results with the OESCN\_a2 and the OESCN\_a1 are shown in Table \ref{table:label1}, the average accuracies of OESCN\_a2, OESCN\_a1, and OESCN are 94.3${\rm{\% }}$, 95.9${\rm{\% }}$, and 97.1${\rm{\% }}$ respectively and the corresponding inter-subject average std of accuracy for OESCN\_a2, OESCN\_a1, and OESCN are 5.3, 4.5, 3.7 respectively. The intra-subject standard deviations are also shown in Table \ref{table:label1}. We observe that both the frequency band attention mechanism and the frequency band generator are effective to improve the classification accuracy, inter-subject robustness, and intra-subject robustness.

\begin{figure}[t]
	\centering
    \includegraphics[width=0.5\textwidth]{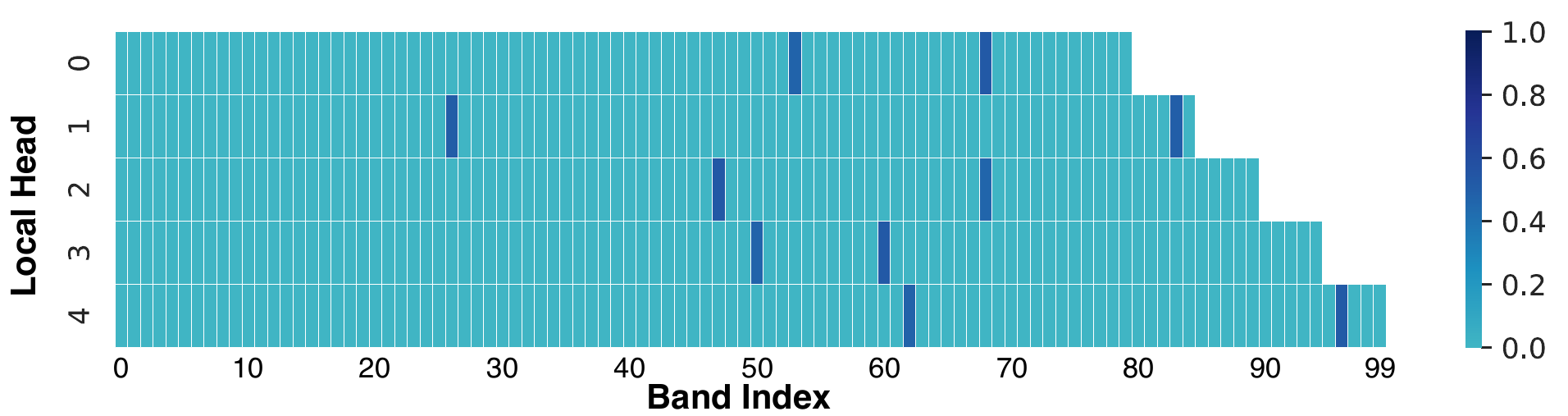}
	\centering
    \includegraphics[width=0.5\textwidth]{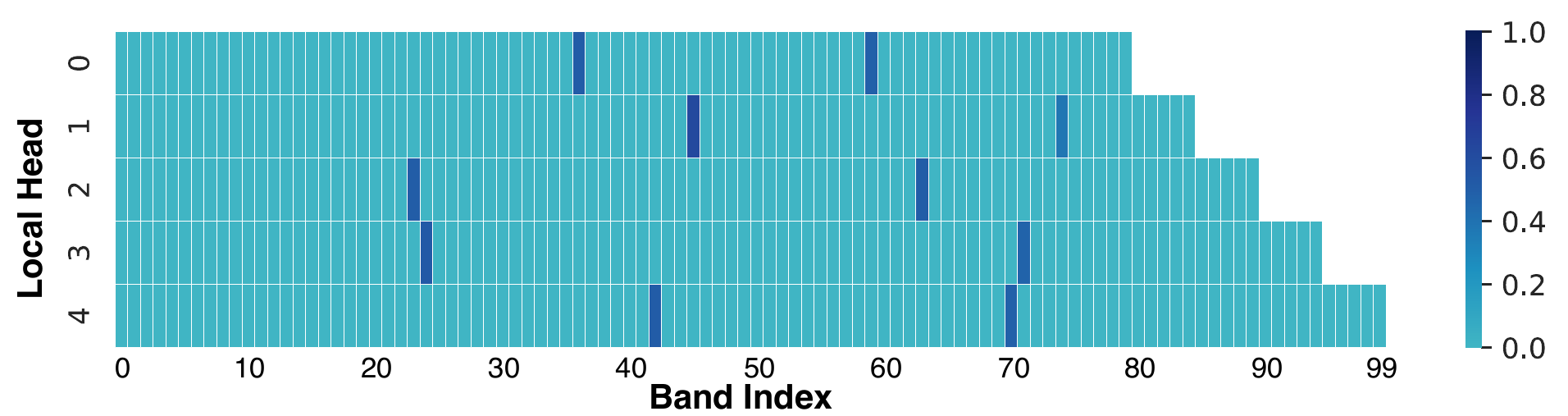}
	\caption{Attention weights of five local head from subject 1 (top) and subject 3 (bottom).}
	\label{fig:Fig. 4}
\end{figure}

We save the attention weights of five local heads from two subjects and draw the heatmap respectively, as shown in Fig. \ref{fig:Fig. 4}. Results show that for each subject the different local heads attend different frequency bands, and generally frequency bands containing high spectral-domain information are more attended. Previous research has shown that high spectral-domain information is highly associated with olfactory EEG  \cite{aydemir2017olfactory}. In addition, the results also show that the attention weights of the two subjects are different.

\section{CONCLUSION}
This letter proposes an olfactory EEG signal classification network (OESCN) based on frequency band feature extraction, which is focused to exploit the spectral-domain information in olfactory EEG signals fully. The experiment results show that the proposed method has high classification accuracy and robustness to inter-subject and intra-subject variations. The proposed approach is not limited to the PSD feature and shows generality, which is also applicable to other spectral-domain features.

\bibliographystyle{IEEEtran}
\balance
\footnotesize
\bibliography{ref}

\end{document}